\documentstyle[aps,prb]{revtex}

\begin{document}






\title{Theta-terms in non-linear
sigma-models}

\author{A. G. Abanov}
\address{12-105, Department of Physics, MIT, 77 Massachusetts Ave.,
Cambridge, MA 02139}
\author{P. B. Wiegmann}
\address{James Franck Institute
and Enrico Fermi Institute of the University of Chicago,\\ 5640 S. Ellis
Avenue, Chicago, IL 60637, USA \\ and Landau Institute for Theoretical
Physics}

\date{\today}
\maketitle

\begin{abstract}
We trace the origin of $\theta$-terms in non-linear $\sigma$-models as
a nonperturbative anomaly of current algebras. The non-linear
$\sigma$-models emerge as a low energy limit of  fermionic
$\sigma$-models. The latter describe Dirac fermions coupled to chiral
bosonic fields. We discuss the geometric phases in three hierarchies
of fermionic $\sigma$-models in spacetime dimension $(d+1)$ with
chiral bosonic fields taking values on $d$-, $d+1$-, and
$d+2$-dimensional spheres. The geometric phases in the first two
hierarchies are $\theta$-terms. We emphasize a relation between
$\theta$-terms and quantum numbers of solitons.
\end{abstract}



\section{Introduction}
 \label{Intro}

Non-linear $\sigma$-models describe the low energy dynamics of
Goldstone bosons emerging as a result of a symmetry breaking. One
particularly important source of this phenomenon is a chiral symmetry
breaking due to an interaction between Dirac fermions and a chiral
bosonic field. This interaction is described by models of current
algebras,  fermionic $\sigma$-models.  Examples are
considered below (\ref{1}-\ref{3},\ref{01}-\ref{21}).

In case of chiral symmetry breaking, the non-linear $\sigma$-models are
determined not only by the pattern of symmetry breaking, but also by
global characteristics of a configurational space of chiral fields. In
most interesting cases configurational space is not simply connected.
There are spacetime configurations of a chiral field which can not be
continuously deformed one into another. An adiabatic motion along a
noncontractible closed path in a configurational space leads to a
geometric phase acquired by the wave function. In the action of a
non-linear $\sigma$-model, the geometric phase emerges as a
$\theta$-term. It is a topological invariant of a mapping of a
spacetime manifold $M$ into a target space $G$ of a chiral
field. Although a $\theta$-term does not appear in equations of
motions, its dramatic impact on dynamics of Goldstone bosons is
needless to emphasize.

The $\theta$-terms reflect the anomalous character of the chiral
current algebra.  Some aspects of chiral anomalies can be studied
perturbatively. For instance, a fermionic current, induced by a
soliton can be obtained in a regular gradient expansion
\cite{GoldstoneWilczek-1981}. The $\theta$-terms are more
involved. They defy a perturbative analysis and are often referred to
as a global (nonperturbative) anomaly.

The $\theta$-terms appear in different physical situations. A
particularly interesting one is when a noncontractible spacetime
configuration is a world trajectory of a soliton carrying a fermionic
number. In this case a $\theta$-term is linked to quantum numbers of
solitons. It is assumed that all $d$ spatial dimensions are
compactified to a sphere $M=S^d$ and the homotopy group
$\pi(M,G)=\pi_d\,(G)\ne 0$ is not zero. Then solitons correspond to
the homotopy classes of the mapping $S^d\to G$. Interaction with
fermions induces a fermionic number localized on a soliton.  This
internal quantum number of a soliton converts into rotational quantum
numbers: spin, statistics and isospin. Intuitively it is appealing
that a fermionic number and spin or statistics of a soliton are two
faces of the same phenomenon. If, for instance, a soliton acquires a
unit fermionic number, one expects it to become a spin 1/2 fermion. In
spite of this, the fermionic number and spin or statistics of a
soliton appear differently in the a non-linear $\sigma$-model. The
fermionic number is assigned by the term $NA_\mu J_\mu$, where $A_\mu$
is an abelian gauge field, $J_\mu$ is a topological current of a
soliton, and $N$ is a number of flavors of a current algebra
\cite{GoldstoneWilczek-1981}. In its turn the spin and statistics are
described by a $\theta$-term. If the homotopy group $\pi(M;G)$ is
nonzero, one can consider an adiabatic $2\pi$-rotation of a soliton
around a given axis as an example of a noncontractible spacetime
configuration. The $\theta$-term of the non-linear $\sigma$-model
represents in this case the geometric phase obtained in this process.

In this paper we intend to clarify some aspects of nonperturbative
anomalies of current algebras and $\theta$-terms in non-linear
$\sigma$-models. We consider two hierarchies of fermionic
$\sigma$-models in spatial dimensions $\,d=0,\,1,\,2,\,3,\cdots$ which
generate $\theta$-terms. The target space of the first hierarchy is a
$d$-dimensional sphere $S^d$. It admits solitons
$\pi_d\,(G)=\pi_d\,(S^d)=Z$. The target space of the second hierarchy
is $S^{d+1}$. There are no solitons in this case $\pi_d\,(S^{d+1})=0$.

Most of non-linear $\sigma$-models we discuss have important physical
applications, some of which will be described elsewhere \cite{AW}. We
show how $\theta$-terms emerge as nonperturbative anomalies of fermionic
$\sigma$-models and how they can be obtained from perturbative
anomalies.  In particular, we show that a non-linear $\sigma$-model of
current algebra which supports solitons and has  noncontractible
spacetime pathes always contains a $\theta$-term with a fixed value of
$\theta=N\pi$.

In the next three sections we summarize the results of the paper. We
write the fermionic models  for two hierarchies  in
Sec.\ref{lsm}, and  list the corresponding non-linear $\sigma$-models
obtained  via gradient expansion in Sec.\ref{nlsm}. The generalization
for arbitrary dimensions is discussed in Sec.\ref{smhd}. We sketch the
computations of anomalous terms  for the first and  second hierarchies
in Secs.\ref{t}-\ref{WZ}. The Sec.\ref{concl} is a summary.

\section{Fermionic $\sigma$-models}
\label{lsm}

{\bf A.}
The first hierarchy of fermionic $\sigma$-models in spatial dimensions
$d=1,\,2,\,3$ is
\begin{eqnarray}%
    (1+1): \hspace{1.0cm} {\cal L}_1 &=&
    \bar\psi\big(i\hat D+im(\Delta_1+i\gamma_5\Delta_2) \big)\psi,
    \hspace{1.0cm} \Delta_1^2+\Delta_2^2=1,
 \label{1}
 \\
    (2+1): \hspace{1.0cm} {\cal L}_2 &=& \bar\psi \big(i\hat D
    +im\vec n\vec\tau\big)\psi, \hspace{2.9cm} \vec n^2=1,
 \label{2}
 \\
    (3+1): \hspace{1.0cm} {\cal L}_3 &=& \bar\psi \big(i\hat D
    +im(\pi_0+ i\gamma_5\vec\pi\vec\tau) \big)\psi,
    \hspace{1.1cm} \pi_0^2+\vec\pi^2=1.
 \label{3}
\end{eqnarray}%
Here and thereon we use Euclidian formulation and assume a mass to be
positive $m>0$.  A Dirac fermion $\psi$ has a flavor running from 1 to
$N$ and $\hat D=\gamma_\mu(\partial_\mu-iA_\mu)$. In dimensions (2+1)
and (3+1) the fermion is also an $SU(2)$ doublet and $\vec\tau$ are
Pauli matrices acting in isospace.  Fermions interact with chiral
fields which take values on spheres $S^d$ (target space).  Hereafter
we reserve a different (historically motivated) notations for $2$-,
$3$-, and $4$-dimensional unit vectors. In one spatial dimension the
chiral field is a phase $\Delta_1+i\Delta_2=e^{i\phi}$. In $d=2$ the
chiral field is a 3-dimensional unit vector $\vec
n=(n_1,\,n_2,\,n_3),\;\;\vec n^2=1$. Correspondingly the chiral field
in three spatial dimensions is a 4-dimensional unit vector
$(\vec\pi,\,\pi_0)$.  It can also be considered as an element of
$SU(2)\; (\approx S^3)$ group $g=\pi_0+i\vec\pi\vec\tau$.

Physical applications of fernionic models (\ref{1}-\ref{3}) on
spheres are known.  In $d=1$ it is familiar Peierls-Fr\"ohlich model
of polyacetylene \cite{Frohlich-1954}. The model in $d=3$ appears in
the past as a model for nuclear forces \cite{TJZW-1985} and recently
has been used in the QCD context \cite{Diakonov-1997}. All these
models also emerged in the context of topological superfluids
\cite{AW}.

In every model of this hierarchy
a compactified coordinate space $S^d$ matches the target space of chiral
fields, also $S^d$. Therefore, every model supports solitons with
an arbitrary \,(integer)\, topological charge $Q$.
The topological
charge $Q$ labels homotopy classes $\pi_d\,(S^d)=Z$ and can be written as a
spatial integral $Q=\int d^dx\,J_0$ of a zeroth component of a topological
current $J_\mu$. In dimensions $d=1,\,2,\,3\,$ topological currents are:
\begin{eqnarray}%
 \label{eq:curr1}
    J_\mu &=& \frac{1}{2\pi} \epsilon_{\mu\nu}\partial_\nu\phi,
 \\
 \label{eq:curr2}
    J_\mu &=& \frac{1}{8\pi}
    \epsilon_{\mu\nu\lambda} \vec n\cdot
    \partial_\nu\vec n\times\partial_\lambda\vec n,
 \\
 \label{eq:curr3}
    J_\mu &=&
    \frac{1}{12\pi^2}\epsilon_{\mu\nu\lambda\rho}
    \epsilon_{abcd}\pi_a\partial_\nu\pi_b
    \partial_\lambda\pi_c\partial_\rho\pi_d
 \\
    &=& \frac{1}{24\pi^2}
    \epsilon_{\mu\nu\lambda\rho}\mbox{tr}
    (g^{-1}\partial_\nu g)(g^{-1}\partial_\lambda g)
    (g^{-1}\partial_\rho g).
 \nonumber
\end{eqnarray}%
On smooth configurations of chiral fields topological currents are
identically conserved $\partial_\mu J_\mu=0$.  We discuss two
different types of spacetime boundary conditions: a compactification
of the spacetime to a sphere $S^{d+1}$, and more physical one -- a
compactification of the space to a sphere $S^d$ at every moment of
time and periodic (anti-periodic) boundary conditions in Euclidian
time. In other words, $M=S^d\times S^1$. In the latter case a
configurational space is divided into disconnected ``topological
sectors'' characterized by total number of solitons $Q$.

In every model of the first hierarchy there are processes which wrap a
spacetime $M=S^d\times S^1$ over a target space $S^{d}$. We consider
simple examples of such processes. In $d=1$ it is a translation of a
soliton around the spatial ring; in $d>1$ it is e.g., a
$2\pi$-rotation of a soliton around a chosen axis.  These spacetime
configurations belong to a nontrivial homotopy classes of $\pi(M;G)$.
These classes can be labeled by an integer $H_d$ (in addition to $Q$),
which for our simple processes appears to be equal to the topological
charge of the soliton $H_d=Q$.

If the spacetime is compactified into a sphere $S^{d+1}$, a
configuration with a given number of solitons at any time does not
exist. Noncontractible spacetime configurations are more
complicated. They consist, e.g., of the creation of a
soliton-antisoliton pair, rotation of a soliton around its axis, and
subsequent annihilation of the pair \cite{WilczekZee-1983}. This
process belongs to a nontrivial homotopy class of
$\pi_{d+1}\,(S^d)$. The corresponding homotopy groups are
$\pi_{2}\,(S^{1})=0$, $\pi_{3}\,(S^{2})=Z$, and
$\pi_{d+1}\,(S^{d})=Z_2$ for $d>2$.

{\bf B.}
Next we consider the second hierarchy of fermionic
$\sigma$-models with the target space $S^{d+1}$. It starts from the
model in $d=0$ (quantum mechanics):
\begin{eqnarray}%
    (0+1):\hspace{1.0cm} { L}_0 &=&
    \bar\psi\big(i\hat D+im(\tau_3\cos\nu
    + \Delta_i\tau_i\sin\nu) \big)\psi,
    \hspace{0.5cm} \Delta_i^2=1,
 \label{01}
 \\
    (1+1):\hspace{1.0cm} { L}_1 &=&
    \bar\psi\big(i\hat D
    +im(\cos\nu+i\gamma_5\vec n\vec\tau\sin\nu) \big)
    \psi, \hspace{0.7cm} \vec n^2=1,
 \label{11}
 \\
    (2+1):\hspace{1.0cm} { L}_2 &=&
    \bar\psi\big(i\hat D+im(\cos\nu\Gamma_5
    + \pi_i\Gamma_i\sin\nu) \big)\psi,
    \hspace{0.5cm} \pi_i^2=1.
 \label{21}
\end{eqnarray}%
Here $\nu$ is some constant, $\Gamma_i$ and $\Gamma_5$ are $4\times 4$
Dirac matrices: $\{\Gamma_i,\,\Gamma_j\}=2\delta_{ij},\;\;\;
\Gamma_5=-\Gamma_1\Gamma_2\Gamma_3\Gamma_4$. Fermions in $d=0,\,1$ are
2-component isospinors of $SO(3)$ and in $d=2$ are 4-component
isospinors of $SO(4)$. The chiral fields do not have solitons
$\pi_d\,(S^{d+1})=0$, but do have noncontractible spacetime
configurations according to the homotopy classes
$\pi_{d+1}(S^{d+1})=Z$.

\section{Non-linear $\sigma$-models}
\label{nlsm}

An action for a chiral field which appears as a result of an
integration over fermions is a non-linear $\sigma$-model ${\cal W}_d
=-\ln \int \exp {(-\int dx{\cal L}_d})D\psi D\bar\psi.\;$ The
non-linear $\sigma$-models can be systematically studied in $1/m\;$
(gradient) expansion. Then anomalies are given by dimensionless terms
of the action of a non-linear $\sigma$-model. They give the only
contribution to the imaginary part of Euclidian action.

Below we list the leading terms of non-linear $\sigma$-models for the
fermionic models of the Sec.\ref{lsm}. They are the main result of
the paper. We start from the first hierarchy (\ref{1}-\ref{3}).

{\bf A.}
In (1+1):
\begin{eqnarray}%
    {\cal W}_1 &=& -N\ln\mbox{Det}\big(i\hat D
    +ime^{i\gamma_5\phi}\big)
 \nonumber \\
    &=& iN\int d^2x A_\mu J_\mu+i\pi N H_1
    +N\int d^2x\frac{1}{8\pi}(\partial_\mu\phi )^2.
 \label{12}
\end{eqnarray}%
Hereafter we omit higher order terms in $1/m$.

The geometric phase $\pi NH_1$ is a $\theta$-term with $\theta=N\pi$. If
the spacetime is compactified to a sphere $S^2$, the geometric phase
vanishes unless $\phi$ is singular (spacetime vortices).
For the spacetime compactified into a torus $T^2=S^1\times S^1$, we have
$\pi(T^2,S^1)=Z\times Z$. There are two integer numbers to characterize
spacetime configurations. One of them is a topological charge $Q=\oint dx\,
\frac{\partial_x\phi}{2\pi}$ while the other is a temporal winding number
$\oint dt\, \frac{\partial_t\phi}{2\pi}$. The geometric phase in this case
is nontrivial even for non-singular configurations. It is a product of
1-cycles over space and time: %
\begin{equation}
    H_1[\phi]=Q \oint dt\, \frac{\partial_t\phi}{2\pi} .
 \label{H11}
\end{equation}%

In (2+1):
\begin{eqnarray}%
    {\cal W}_2 &=& -N\ln\mbox{Det}\big(i\hat D+im\vec n\vec\tau\big)
 \nonumber\\
    &=& iN\int d^3x A_\mu J_\mu+i\pi NH_2[\vec n]
    + N\frac{m}{8\pi}\int d^3x\,(\partial_\mu\vec n)^2.
 \label{22}
\end{eqnarray}%
Here $\pi N H_2$ is a $\theta$-term with $\theta=\pi N$. The integer
$H_2$ is a homotopy class of a map of the spacetime to the target
space $S^2$. In the case when the spacetime is compactified to a
sphere $S^3$, $H_2$ is the Hopf number. It can be any integer
$\pi_3\,(S^2)=Z$. A geometric interpretation of the Hopf number is
well known (see e.g., Refs.  \cite{WilczekZee-1983,DFN-1985}). It is a
linking number of world lines of two different values $\vec n_1$ and
$\vec n_2$ of $\vec n$-field. The Hopf number can be explicitly
written in terms of $SU(2)$ matrix $U(x)$ which rotates the vector
$\vec n$ to the chosen (third) axis: $\vec n\vec\tau=U^{-1}\tau_3
U$. Then the Hopf number is a degree of mapping of (2+1) spacetime
into $SU(2)$:
\begin{equation}
    H_2[\vec n] = \frac{\epsilon^{\mu\nu\lambda}}{24\pi^2}
    \int d^3 x\, \mbox{tr}(U^{-1}\partial_\mu U)\,
    (U^{-1}\partial_\nu U)\,(U^{-1}\partial_\lambda U).
 \label{H22}
\end{equation}

If the spacetime is compactified into $M=S^2\times S^1$, the topology
of the mapping $M\to S^2$ is more complicated. The homotopy classes
are labeled by two integers $(Q, H_2)$. The first is a topological
charge of solitons. It is the same at all times. The second integer
$H_2$ is a topological invariant of spacetime configurations. $H_2$ is
generalizing the Hopf invariant, defined for a map $S^3\to S^2$.

In general, the homotopy classes of maps $S^2\times S^1\to S^2$ do not
form a group.  However, they do form a group in a sector with a fixed
topological charge.  In a sector with zero topological charge and if
the vector $\vec n(x=\infty,t)$ on the space infinity does not depend
on time, $H_2$ is the Hopf invariant. It is a linking number of world
trajectories of two arbitrarily chosen points $\vec n_1$ and $\vec
n_2$ of the target space. However, in a sector with non-zero
topological charge some links can be unlinked with the help of a
continuous deformation of the vector at space infinity. To construct a
proper topological invariant one has to consider the angle of rotation
of $\vec n\,(x=\infty,t)$ around one of the vectors $\vec n_1$ or
$\vec n_2$ as it is illustrated in Ref. \cite{MM-1995}. The new
invariant, however, is intrinsically ambiguous and is defined modulo
$2Q$ only. There is a deep mathematical theorem, which says that in a
sector with topological charge $Q$ there are only $2Q$ homotopy
classes, and that they form a finite Abelian group $Z_{2Q}\,$
\cite{Pontrjagin-1941three}. In a sector with $Q$ solitons a $2\pi k$
rotation of a single soliton with $k=0,\cdots,2Q-1$ is a
representative spacetime configuration of the $k$-th homotopy
class. If $\phi(t)$ is an angle of rotation of a soliton of charge $Q$
around fixed axis, the invariant $H_{2}$ is given by the same formula
(\ref{H11}) as in (1+1) case. It is a product of the spatial 2-cycle
$Q$ and the temporal 1-cycle. A more detailed discussion is planned
elsewhere \cite{AW}.

Finally in (3+1):
\begin{eqnarray}%
    {\cal W}_3 &=& -N\ln\mbox{Det}\big(i\hat D
    +img^{\gamma_5}\big)
 \nonumber\\
    &=& iN\int d^3x A_\mu J_\mu+i\pi N H_3[g]
    +N\frac{F_\pi^2}{4} \int d^3x\,
    \mbox{tr}(\partial_{\mu}g^{-1}\partial_\mu g),
 \label{32}
\end{eqnarray}%
where $F_\pi^2=\frac{1}{2\pi^2}m^2\ln\frac{\Lambda}{m}$ with $\Lambda$
being ultraviolet cutoff and we use notation
$g^{\gamma_5}=\frac{1+\gamma_5}{2}g +\frac{1-\gamma_5}{2}g^{-1}$. The
integer $H_3$ is a topological invariant of the map of the spacetime
into the target space $S^3$. There are only two homotopy classes
$\,\pi_4\,(S^3)=Z_2$, so that $H_3=0$ or $1$. Geometric phase $\pi N
H_3$ is the $\theta$-term with $\theta=N\pi$.

In general, the finite number of homotopy classes gives a restriction
for admissible values of $\theta$. The wave function or $e^{i{\cal
W}}$ forms a single-valued representation of the homotopy group only
if $\theta$ is a multiple of $2\pi/l$, where $l$ is a dimension of
the homotopy group. In $(3+1)$ and higher dimensional models of this
hierarchy $\,\pi_{d+1}\,(S^d)=Z_2$. This limits the value of $\theta$
to multiples of $\pi$.

This restriction is of a particular interest in (2+1) case. In a
sector with $Q$ solitons the allowed values of $\theta$ are multiples
of $\pi/Q$. Even in the case when the spacetime is compactified to a
sphere, where there is no formal restriction on a value of $\theta\,$
\cite{WilczekZee-1983}, solitons and antisolitons can not be treated
as true particles, unless $\theta$ is quantized in units of
$\pi/Q$. If one does not want to restrict a $\sigma$-model to a sector
with a given number of solitons, the only allowed values of $\theta$
are multiples of $\pi$ (see also Ref. \cite{GovShan-1989}).

{\bf B.}
Let us now list the non-linear $\sigma$-models of the second hierarchy
(\ref{01}-\ref{21}).
\begin{eqnarray}%
    (0+1):\hspace{1.0cm} { W}_0
    &=& N\int dt\, \frac{\sin^2\nu}{8m}(\partial_t\phi)^2
    -i\theta \Omega_1\left[\phi \right],
 \label{210}
 \\
    \Omega_1\left[\phi \right] &=&
    \oint \,  \partial_t\phi \frac{dt}{2\pi}.
 \nonumber
\end{eqnarray}%
This is the Lagrangian of a plane quantum rotator moving around magnetic
flux $\theta=\pi N(1-\cos\nu)$ i.e., (0+1)-dimensional $O(2)$ non-linear
$\sigma$-model with $\theta$-term.
\begin{eqnarray}
    (1+1):\hspace{1.0cm} { W}_1 &=&
    N\int d^2x\,\frac{\sin^2\nu}{4\pi}(\partial_\mu\vec n)^2 -
    i\theta\Omega_2\left[\vec{n} \right],
 \label{211}
 \\
    \Omega_2\left[\vec{n} \right] &=& \int d^2x\,
    \frac{1}{8\pi} \epsilon_{\mu\nu\lambda}
    \epsilon_{abcd} n_a\partial_\nu n_b\partial_\lambda n_c .
 \nonumber
\end{eqnarray}
This is the famous $(1+1)$, $O(3)$ non-linear $\sigma$ model with
$\theta =N(2\nu-\sin2\nu)$. For $\nu=\pi/2$ we find $\theta=N\pi$ and
this is the non-linear $\sigma$-model with topological term which was
used to describe the effective action for spin-$N/2$ chain
\cite{Haldane-1983PLA}.
\begin{eqnarray}%
    (2+1):\hspace{1.0cm} { W}_2
    &=& N\int d^3x\,\frac{m\sin^2\nu}{4\pi}
    (\partial_\mu \pi_i)^2 -i\theta \Omega_3\left[\pi \right],
 \label{212}
 \\
    \Omega_3\left[\pi \right] &=& \int d^3x\,
    \frac{1}{12\pi^2}\epsilon^{\mu\nu\lambda}\epsilon^{ijkl}
    \pi_i\partial_\mu \pi_j
    \partial_\nu \pi_k\partial_\lambda \pi_l.
 \nonumber
\end{eqnarray}%
This is a $(2+1)$, $O(4)$ non-linear $\sigma$-model  with
$\theta=N\pi(1-\frac{3}{2}\cos\nu+\frac{1}{2}\cos 3\nu)$.

The anomalous part $A_\mu J_\mu$ in (\ref{12},\ref{22},\ref{32}) with
$J_\mu$ from (\ref{eq:curr1}-\ref{eq:curr3}) has been computed in
Ref. \cite{Frohlich-1954,Jaroszewicz-1984,GoldstoneWilczek-1981} for
$d=1,\,2,\,3$ respectively. The geometric phase in (1+1) on a
spacetime compactified into a torus ({\ref{H11}) appeared in a context
of bosonization on general Riemann surfaces (see
Ref. \cite{ABMNV-1986}).  The geometric phase of (\ref{22}) has been
considered in Ref. \cite{Jaroszewicz}. The geometric phase in (3+1) is
related to the so-called $SU(2)$ anomaly and can be found in
Ref.\cite{Witten-1983baryons}. $O(3)$ non-linear $\sigma$-model at
$\nu=\pi/2$ has been obtained from (1+1) fermionic model
(\ref{11}) in Ref. \cite{Tsvelik-1994}.

Below we also mention one more (third) hierarchy, with the target space
$S^{d+2}$.  These models do not have solitons $\pi_d\,(S^{d+2})=0$,
and they do not have $\theta$-terms either: $\pi_{d+1}(S^{d+2})=0$. Their
geometric phases are perturbative Wess-Zumino terms due to
$\pi_{d+2}\,(S^{d+2})=Z$.

\section{ $\sigma$-models in higher dimensions} \label{smhd}

Obviously both hierarchies may be continued to higher dimensions.
Their topological properties are ``inherited'' from the ones in lower
dimensions due to the stability of homotopy groups \cite{stability}:
$\pi_{n+k}(S^n)$ are the same for all $n\ge k+2$. To write the models
explicitly we introduce the following notations.  Let
$\vec{V}=(V_1,\cdots,V_l)$ be a a unit vector $\sum_i V_i^2=1$ which
takes values on $(l-1)$-sphere $S^{l-1}$ and
$\Gamma_1^{(2k+1)},\cdots,\Gamma_{2k+1}^{(2k+1)}$ are $2^k\times 2^k$
Hermitian Dirac matrices (representation of Clifford algebra with
$2k+1$ generators). We denote
\begin{eqnarray}
 \label{87}
    V^{(l)} &=& \sum_{i=1}^l
    V_i\Gamma_i^{(l)}; \hspace{3.0cm} l=\mbox{odd} ,
 \\
 \label{88}
    V^{(l)} &=& V_l+i\gamma_5\sum_{i=1}^{l-1}
    V_i\Gamma_i^{(l-1)};
    \hspace{1.0cm} l=\mbox{even} .
\end{eqnarray}
Here $V^{(l)}$ is a matrix with unit square
$V^{(l)}(V^{(l)})^\dagger=1$.  In these notations the hierarchy of
fermionic models (\ref{1}-\ref{3}) with the target space $S^d$
in $d+1$ dimensions is
\begin{equation}
    {\cal L}_d =\bar\psi(i\hat D
    +im V^{(d+1)})\psi .
 \label{81}
\end{equation}
The non-linear $\sigma$-model for this hierarchy is
\begin{equation}
 \label{83}
    {\cal W}_{d}= iN\int d^{d+1}x\,
    A_\mu J_\mu +i\pi N H_d
    + N m^{d-1}\alpha_{d}\int d^{d+1}x\,
    (\partial_\mu \vec{V})^2,
\end{equation}
where $\alpha_{d}$ is a constant (non-universal for $d>3$,
$\alpha_{d}\sim \left(\frac{\Lambda}{m}\right)^{d-3}$),
$\alpha_{3}=\frac{1}{4\pi^2}\ln\frac{\Lambda}{m}$. Topological
invariant $H_d$ takes values $0$ and $1$ for $d\geq 3$ according to two
homotopy classes $\pi_{d+1}(S^d)=Z_{2}$  and $J_\mu$ is a
topological current
\begin{equation}
 \label{eq:topcurrd}
    J_\mu = \frac{1}{d!\,\mbox{Area}(S^{d})}
    \epsilon^{\mu\mu_1\ldots\mu_{d}}
    \epsilon^{a a_1\ldots a_{d}}
    V_a \partial_{\mu_1} V_{a_1}\cdots
    \partial_{\mu_{d}} V_{a_{d}}.
\end{equation}
Here
$\mbox{Area}(S^{d})=\frac{2\pi^{\frac{d+1}{2}}}{\Gamma(\frac{d+1}{2})}$ is
an area of unit $d$-sphere. Zero component of topological current
integrated over space is a topological charge of configuration $Q=\int
d^dx\, J_0$---the degree of mapping $S^d\to S^d$.

The hierarchy of fermionic models \,(\ref{01}-\ref{21}) with
the target space $S^{d+1}$ is given by
\begin{equation}
    L_d =\bar\psi( i\hat D+im V^{(d+3)})\psi
 \label{84}
\end{equation}
with a condition
\begin{equation}
 \label{888}
    V_{d+3}=\cos\nu,\hspace{1.5cm}
    \sum_{i=1}^{d+2}V_i^2=\sin^2\nu.
\end{equation}
If $\vec{v}$ is a unit $d+2$-vector ($\vec{v}^2=1$) defined as
$\vec{v}\sin\nu \equiv (V_1,\ldots,V_{d+2})$ and matrix $v$ is defined
as $v\equiv \sum_{i=1}^{d+2}v_i\Gamma_i$, then this hierarchy can also
be written as
\begin{eqnarray}
 \label{848}
    L_d &=& \bar\psi\Big(i\hat D
    +im (\cos\nu\Gamma_{d+3}+\sin\nu\, v)\Big)\psi;
    \hspace{1.5cm} d+1=\mbox{odd},
 \\
 \label{849}
    L_d &=& \bar\psi(i\hat D+im e^{i\gamma_5 \nu v})\psi;
    \hspace{4.1cm} d+1=\mbox{even}.
\end{eqnarray}
The non-linear $\sigma$-model for this hierarchy is
\begin{equation}
 \label{877}
    W_d= -i\theta\Omega_d
    +N m^{d-1}2\alpha_{d}\sin^2\nu
    \int d^{d+1}x\, (\partial_\mu \vec{v})^2 .
\end{equation}
The $\theta$-term $\Omega_d$ is a degree of mapping $S^{d+1}\to S^{d+1}$:
\begin{equation}
 \label{eq:Odexpr}
    \Omega_d = \frac{1}{(d+1)!\,
    \mbox{Area}(S^{d+1})} \int d^{d+1}x\,
    \epsilon^{\mu_1\ldots\mu_{d+1}}
    \epsilon^{a a_1\ldots a_{d+1}} v_a
    \partial_{\mu_1} v_{a_1}\cdots
    \partial_{\mu_{d+1}} v_{a_{d+1}}
\end{equation}
and coefficient $\theta$ is proportional to a fraction of the area of
$d+2$-sphere $S^{d+2}$ cut off by latitude $\nu$
\begin{equation}
 \label{eq:theta}
    \theta = 2\pi N
    \frac{\int_0^{\nu}\sin^d z \,dz}
    {\int_0^{\pi}\sin^d z \,dz}.
\end{equation}

\section{$\theta$-term and quantum numbers of  soliton}
 \label{t}

We start from the first hierarchy (\ref{1}-\ref{3}). It is well known
that a soliton in these models acquires a fermionic charge due to
a chiral anomaly. It is described by the first term in non-linear
$\sigma$-models (\ref{12},\ref{22},\ref{32}).  Varying over the gauge
field, we obtain a fermionic current induced by a soliton
\begin{equation}
 \label{J}
    j_{\mu}=-i\frac{\delta}{\delta A_{\mu}}{\cal W}_{d}=NJ_\mu.
\end{equation}
This result means that a Hamiltonian of a fermionic model taken in a
static soliton background have an additional $NQ$ levels with negative
energy $E<0$ compared to the one in a topologically flat
background. Computations of the induced current are perturbative and
known \cite{GoldstoneWilczek-1981,Jaroszewicz,NiemiSemenoff-1986}. We
write ${\cal L}=\bar\psi {\cal D}\psi$ and
$j_{\mu}=i\frac{\delta}{\delta A_{\mu}}N \mbox{Tr}\ln{\cal D}
=iN\mbox{Tr}\left\{\gamma_{\mu} ({\cal D}^{\dagger}{\cal D})^{-1}{\cal
D}^{\dagger}\right\}$. Expanding the denominator in gradients of
chiral fields
\begin{equation}
 \label{00}
    j_{\mu} = mN\mbox{Tr}
    \left\{\gamma_{\mu} \frac{1}{-\partial^2+m^2}
    \left(m\hat \partial V^{(d+1)}\frac{1}{-\partial^2+m^2}\right)^{d}
    (V^{(d+1)})^{\dagger}\right\},
\end{equation}
and calculating the trace, we obtain the result (\ref{J}) with
topological current (\ref{eq:topcurrd}). For dimensions $d=1,2,3$ this
gives (\ref{eq:curr1}-\ref{eq:curr3}).  This is the first nonvanishing
term in the gradient expansion for current.  One can use (\ref{00})
only for large size ($\gg 1/m$) solitons. If the soliton is small, the
gradient expansion is not applicable: some levels cross $E=0$ and
soliton becomes uncharged \cite{MKWil-1984ill}.

The leading regular term of effective actions (\ref{12},\ref{22},\ref{32})
can be obtained in a similar fashion by varying over a chiral field:
\begin{eqnarray}
    \delta {\cal W}_d &=& -N\mbox{Tr}
    \left\{im\delta V^{(d+1)}
    ({\cal D}^{\dagger}{\cal D})^{-1} {\cal D}^{\dagger}\right\}
 \nonumber \\
    &=& Nm^2\mbox{Tr}
    \left\{\delta V^{(d+1)}\frac{1}{-\partial^2+m^2}
    \hat{\partial}V^{(d+1)}
    \frac{1}{-\partial^2+m^2} \hat{\partial} \right\}.
 \nonumber
\end{eqnarray}

The calculation of the $\theta$-term is more complicated due to its
nonperturbative nature. Below we employ the method suggested in Ref.
\cite{Jaroszewicz}. We illustrate it on the example of (2+1) theory
(\ref{2}) and choose the spacetime to be compactified into
$S^{2}\times S^{1}$. We specify $\vec{n}(\vec
x,t)\vec\tau=e^{-\frac{i}{2}\varphi(t)\tau^{3}}\vec{n}_{0}\vec\tau
e^{\frac{i}{2}\varphi(t)\tau^{3}}$ to be a charge $Q$ soliton slowly
rotating around ``third'' axis given by $\vec n_{0}(\vec x=\infty)$
and performing total $2\pi$-angle rotation
$\varphi(T)-\varphi(0)=2\pi$ in time $T$. Then we compare the value of
the determinant of the Dirac operator
$\mbox{Det}\,(i\hat{\partial}+im\vec n\,(\vec x,t)\vec\tau)$ with the
one for a static, nonrotating soliton $\vec{n}_0(\vec x)$.  According
to the Sec.\ref{nlsm}, we expect that a difference between topological
terms $H_2$ for these two configurations is $Q$. Then the ratio of the
determinants is $e^{i\theta Q}$ which gives the value of the
coefficient $\theta$ in front of $\theta$-term.  Other (regular) terms
are negligible as soon  rotation is adiabatically slow.

To facilitate computation it is tempting to make a gauge
transformation $\psi\rightarrow e^{-\frac{i}{2}\varphi
\tau^{3}}\psi'$. Then the transformed Dirac operator $ i\hat D
+\frac{1}{2}\gamma_{0}\dot{\varphi}\tau^{3} +im\vec{n}_{0}\vec\tau$
depends on time derivative of $\varphi$ and is ready for a gradient
expansion. This approach, however, misses the geometric phase. The
resolution of this puzzle is typical for an anomalous calculus. The
gauge transformation we performed is anomalous. It is not
single-valued: a change of $\varphi$ by $2\pi$ changes the sign of
$\psi$ if $Q$ is odd. This transformation changes the antiperiodic
boundary condition $\psi(0)=-\psi(T)$ to the periodic one $\psi'(0)=
-e^{\frac{i}{2}\oint\dot\varphi dt}\psi'(T)=(-1)^{Q+1}\psi'(T)$ by
creating a flux $\pi$ through the temporal loop. The situation may be
improved by making an additional Abelian gauge transformation $\psi'
\rightarrow e^{\frac{i}{2}\varphi}\psi''$. Then ${\cal D}\to i\hat{D}
+\frac{1}{2}\gamma_{0}\dot{\varphi}\tau^{3}
+\frac{1}{2}\gamma_{0}\dot{\varphi} +im\vec{n}_{0}\vec\tau $. Now
boundary conditions for $\psi''$ are the same as for $\psi$ and we can
expand the transformed Dirac operator in $\dot\varphi$. At this point
we notice that $\frac{1}{2}\dot\varphi$ enters Dirac operator the same
way as constant in space gauge potential $A_{0}$. Then the variation
of an effective action over $\dot\varphi$ can be computed the same
way as it has been done for the calculation of an induced current. The
result is expressed in terms of the induced charge. We have
\begin{eqnarray}
 \label{b}
    -i\frac{\delta {\cal W}_{d}}{\delta \dot\varphi}
    &=& iN\mbox{Tr}\left\{\frac{1+\tau_3}{2}
    \gamma_0 ({\cal D}^{\dagger}
    {\cal D})^{-1}{\cal D}^{\dagger}\right\}
 \nonumber \\
    &=& iN\mbox{Tr}\left\{\frac{1}{2}\gamma_0
    ({\cal D}^{\dagger}{\cal D})^{-1}
    {\cal D}^{\dagger} \right\}
    = -\frac{i}{2}\frac{\delta {\cal W}_{d}}{\delta A_0}
    =\frac{N}{2}\int d^dx\,J_0.
\end{eqnarray}
Thus we obtain an important result: an angular momentum (spin) of a
soliton $I=-\frac{\delta}{\delta \dot\varphi}{\cal W}_{d}$ is equal to
the half of the topological charge of a soliton times the degeneracy
of fermionic states
\begin{equation}
 \label{I}
    I= N\frac{Q}{2}.
\end{equation}
This gives us the topological term $\theta H_2$ in (\ref{22}) with
$\theta=N\pi$.

A computation of the geometric phase for a configuration other than a
rotating soliton is technically involved. It is  not necessary
though due to a topological nature of the geometric phase. It is a
topological invariant and does not change within a homotopy class. We
may, therefore consider a particular configuration---$2\pi$ rotation
of a soliton of charge $Q$ to find a $\theta$-angle, a coefficient in
front of the topological invariant. Generalization of these arguments
to higher dimensions is straightforward.

These arguments clearly relate the fermionic number of a soliton with
its angular momentum (\ref{I}) and, therefore, its statistics. If
solitons are charged they also have nonzero angular momentum (and
corresponding statistics). The latter translates into the
$\theta$-term in a non-linear $\sigma$-model. The value of $\theta$ in
this term is determined by the charge of soliton. Since the fermionic
charge in our models is an integer the value of $\theta$ is always a
multiple of $\pi$. Contrary the $\theta$-terms in the second hierarchy
can get any value.

Despite different dimensions of models belonging to the first
hierarchy, statistical properties of solitons do not depend on a
dimension of the spacetime. Here is a brief list of them.

(i) A soliton with a topological charge $Q$ carries a fermionic number $NQ$.

(ii) A soliton of unit topological charge in a model with odd (even)
number of flavors $N$ has a half-integer (integer) spin. To see this
in spatial dimensions 2 and 3 in a spacetime compactified into a
sphere, one might consider an adiabatic process when
soliton-antisoliton pair is created, the soliton is rotated by
$2\pi$-angle, and then the pair is annihilated. This process
corresponds to a spacetime configuration from nontrivial homotopy
class. The value of $\theta$-term in the action for this spacetime
configuration is $N\pi$ and the wave function of the entire system changes
its sign under $2\pi$-rotation.  We conclude that this soliton has a
spin $N/2$.  We notice that although the $\theta$-term vanishes modulo
$2\pi$ at even $N$ its effect does not disappear. A soliton in this
case has an integer spin
\cite{Witten-1983global,WilczekZee-1983,BNRS-1982} equal to $N/2$.

(iii) A process when two solitons interchange their spatial positions
also changes the topological invariant by one\cite{WilczekZee-1983}
and the geometric phase by $N\pi$. This leads to a conclusion that a
soliton is a fermion (boson) in case $N$ is odd (even).

(iv) In the toroidal space geometry (spatial dimensions are compactified
into a torus) the $\theta$-term changes momentum quantization rule
from $P=\frac{2\pi}{L}n$ to $P=\frac{2\pi}{L}(n+\frac{NQ}{2})$ in the
sector with $Q$ solitons. This change is nontrivial if $NQ$ is odd.

(v) The $\theta$-term can be also interpreted locally as a Berry
phase \cite{NiemiSemenoff-1985} of an adiabatically rotating soliton.
The total phase acquired by the vacuum state as a result of
$2\pi$-rotation is the $\theta$-term.

\section{Wess-Zumino and $\theta$-terms}
 \label{WZ}

Noncontractible spacetime configurations of the second hierarchy
(\ref{01}-\ref{21},\ref{84}) are not linked to any particle-like
states.  The analysis of $\theta$-terms in this case is
simpler. They may be obtained through a reduction from the Wess-Zumino
perturbative anomaly \cite{Witten-1983baryons}. The construction is
following. Let us increase the target space $S^{d+1}$ of the models
(\ref{01}-\ref{21},\ref{84}) to $S^{d+2}$. Then all spacetime
configurations become contractible $\pi_{d+1}\,(S^{d+2})=0$. It does
not mean, however, that a geometric phase vanishes. It exists due to a
nonzero homotopy group $\pi_{d+2}(S^{d+2})=Z$ (See
Ref. \cite{Witten-1983global}). In this case the geometric phase is
perturbative and is known as Wess-Zumino term. Under the reduction
back to $S^{d+1}$, the Wess-Zumino term converts into a nonperturbative
$\theta$-term. To carry out this procedure, we consider the third
hierarchy of fermionic models with the target space $S^{d+2}$.  They
are the same as (\ref{84}) but with no constraint (\ref{888}) (a
constant $\nu$ becomes a dynamic field):
\begin{equation}
 \label{LL}
    {\cal D}=i\hat D+imV,
\end{equation}
where we adopt the notation $V^{(d+3)}=V$.

In dimensions $d=0,\,1$ these are familiar models:
\begin{eqnarray}
 \label{71}
    (0+1):\hspace{1.0cm} {\cal D} &=& i D+im\vec n\vec\tau ,
 \\
 \label{72}
    (1+1):\hspace{1.0cm} {\cal D} &=& i\hat
    D+im(\pi_0+i\gamma_5\vec \pi\vec\tau).
\end{eqnarray}

We shall compute the perturbative anomaly for the third hierarchy
(\ref{LL}) and then enforce the condition (\ref{888}).

Varying over $V$, we have $\delta W=-N\mbox{Tr}\,\Big(im\delta V
(D^\dagger D)^{-1}D^\dagger\Big)$. Expanding $ {{\cal (D^\dagger
D)}}^{-1}$ in gradients
of $V$ we obtain for an imaginary part
\begin{equation}
 \label{55}
    \delta \Im m W =iNK \int_{S^{d+1}} d^{d+1}x\,
    \mbox{Tr}\left(\delta V (\hat\partial V)^{d}V^\dagger\right).
\end{equation}
Here
$K=\int\frac{d^{d+1}p}{(2\pi)^{d+1}}\,\frac{m^{d+3}}{(p^2+m^2)^{d+2}}$
and trace is taken over both Lorentzian and isospace gamma-matrices.
Integration is performed over the spacetime $S^{d+1}$.
The method to restore the action of non-linear $\sigma$-model is
standard \cite{Witten-1983global}. Introduce a parameter $\xi$ such that
$\vec{V}(x,\xi)$
continuously interpolates between constant $\vec{V}(x,\xi=0)=
(\vec{0},V_{d+3})$ and given
spacetime configuration $\vec{V}(x,\xi=1)= \vec{V}(x)$.
The field $\vec{V}(x,\xi)$ is
therefore defined on a disk $\cal{B}$, which boundary
$\partial {\cal B}=S^{d+1}$ is the spacetime of
our model. Take
a Jacobian of a map of $d+2$-sphere $(x,\xi)$ to $d+2$-sphere
$\vec{V}(x,\xi)$ and
integrate it over the disk:
\begin{eqnarray}
 \label{56}
    \Im m W &=& -2\pi N\Gamma [\vec{V}],
 \\
    \Gamma [\vec{V}] &\equiv &
    \frac{1}{(d+2)!\mbox{ Area}(S^{d+2})}
    \int_{\cal B} d^{d+2}x\,
    \epsilon^{\mu_1\ldots\mu_{d+2}}
    \epsilon^{a a_1\ldots a_{d+2}} V_a
    \partial_{\mu_1} V_{a_1}\cdots
    \partial_{\mu_{d+2}} V_{a_{d+2}}.
 \nonumber
\end{eqnarray}
Since the integrand is a full derivative the result of the
integration, depends only on the physical field defined on the
boundary of the disk. Its variation gives (\ref{55}). Adding a leading
term of the gradient expansion of the real part of $-N\mbox{Tr}\ln\cal
D$ we obtain the non-linear $\sigma$-models for the third hierarchy
(\ref{LL}):
\begin{equation}
    W=\frac{F_\pi^2}{2}\int d^{d+1}x\,
    (\partial_\mu \vec{V})^2-2\pi iN\Gamma [\vec{V}].
\end{equation}
These are Wess-Zumino models.
In (0+1) it is an action for a spin $N/2$ (see e.g.,
\cite{Stone-1986,Wiegmann-1989})
\begin{equation}
 \label{91}
    W_0=\frac{N}{8m}
    \int dt\, (\dot{\vec n})^2-2\pi iN\int d^2x\,
    \frac{1}{8\pi}\epsilon^{\mu\nu}
    \vec n\cdot\partial_{\mu}\vec n\times\partial_\nu\vec n.
\end{equation}
In (1+1) it is $SU(2)$ level $N$ WZ-model of conformal field theory
\cite{PolWieg-1983,Witten-1984bos,KnizhnikZam-1984}:
\begin{eqnarray}
 \label{92}
    W_1 &=& \frac{N}{8\pi}\int d^2x\,
    \mbox{Tr}\,(\partial_\mu g^{-1}\partial_\mu g)
 \\
    &-& 2\pi iN\int d^3x\,
    \frac{1}{24\pi^2}\epsilon^{\mu\nu\lambda}\mbox{Tr}
    (g^{-1}\partial_\mu g)(g^{-1}\partial_\nu g)
    (g^{-1}\partial_\lambda g).
 \nonumber
\end{eqnarray}

Now let us perform a reduction of the target space $(d+2)$-sphere of
the Wess-Zumino models to $S^{d+1}$ by imposing constraint
(\ref{888}). It brings us back to the models of interest
(\ref{210}-\ref{212},\ref{877}-\ref{eq:theta}). The condition (\ref{888})
embeds a $(d+1)$-sphere into a $(d+2)$-sphere as a latitude $\nu$
section.  Under this condition, $\vec{V}$ on the boundary of the disk
$(x,\xi=1)$ (physical spacetime) takes values on $S^{d+1}$. In this
case the Wess-Zumino term is equal to a degree of a mapping
$S^{d+1}\to S^{d+1}$ times the fraction of the volume of $S^{d+2}$ cut
off by the latitude corresponding to the angle $\nu$. This factor is
the value of $\theta$ in the models
(\ref{210}-\ref{212},\ref{877}-\ref{eq:theta}). It is given by
(\ref{eq:theta}).  In particular at $\nu=\pi/2$ the value of the
$\theta$-angle is $N\pi$.

\section{Summary}
\label{concl}

To summarize the results of the paper: we discussed connections
between global properties of chiral current algebras and geometric
phases in non-linear $\sigma$-models.  To understand these connections
the mappings of three different manifolds to the manifold (target
space) of the chiral field are essential. These manifolds are: (i)
space, (ii) spacetime, and (iii) a disk which
boundary is a spacetime.

We considered three hierarchies of models of Dirac fermions in
$(d+1)$-dimension coupled with chiral boson field. The chiral fields
take values on $d$-, $d+1$-, and $d+2$-dimensional spheres
respectively. Each hierarchy provides a different physical origin of a
geometric phase. Most of the models in dimensions $d=1,2,3$ have
important physical applications.

The current algebras with a $d$-dimensional sphere as a target space
(first hierarchy) support solitons---nontrivial homotopy classes of
spatial configurations of the chiral field. Solitons carry an integer
fermionic charge. We showed (see also Ref. \cite{Jaroszewicz}) that
the induced actions for the chiral field (non-linear $\sigma$-models)
necessarily possess $\theta$-terms. These terms represent homotopy
classes of $d+1$-dimensional spacetime configurations. The value of
$\theta$-angle is tightly related to the charge of solitons. It is
equal to $\pi$ times the fermionic number of the soliton with a unit
topological charge. The $\theta$-term is a geometric phase reflecting
a spin and statistics of the soliton. The spin appears to be equal a
half of the fermionic number, thus establishing a relation between a
fermionic number and spin (statistics) of a soliton.

The current algebra with $S^{d+1}$ target space (the second hierarchy)
does not have solitons. All spatial configurations are
contractible. However, spacetime configurations are not. As a result
there is a $\theta$-term in the non-linear $\sigma$-model. Contrary to
the first hierarchy, a $\theta$-angle is not restricted and may take
any value, depending on the value of the parameter in the fermionic
model.  We obtained it as a result of a reduction from current
algebras on ${d+2}$-spheres.

Finally for the current algebras on $d+2$-dimensional spheres (the third
hierarchy) both spatial and spacetime configurations are
contractible. As a result, there is no
$\theta$-term in the non-linear $\sigma$-model. However, the
geometric phase exists due to nontrivial extensions of spacetime
configurations to a $d+2$-dimensional sphere. This geometric phase
is the WZ-term with a coefficient equal to the number of
flavors of fermions. The non-linear $\sigma$-models in this case can be
obtained by a regular gradient expansion.

\section{Acknowledgment}

We would like to thank S. P. Novikov and H. R. Miller for discussing
topology. A. A.  is grateful to P. A. Lee, X.-G. Wen, and D. A. Ivanov
for many fruitful discussions. P. W.  would like to thank the Lady
Davis foundation for the hospitality in Hebrew University in
Jerusalem, where this work has been completed. A. A. was supported by
NSF DMR 9813764. P.  W. was supported by grants NSF DMR 9971332 and
MRSEC NSF DMR 9808595.


\begin{thebibliography}{99}



\bibitem{GoldstoneWilczek-1981}J. Goldstone and F. Wilczek, Phys. Rev.
Lett. {\bf 47}, 986-989 (1981).
\\ {\it Fractional Quantum Numbers on Solitons}

\bibitem{AW} A. G. Abanov and P. B. Wiegmann, to be published.

\bibitem{Frohlich-1954}
G. Fr\"{o}hlich, Proc. R. Soc. {\bf A223}, 296-305 (1954),
\\ {\it On the theory of superconductivity:
the one-dimensional case} \\
For review see:
S. A. Brazovskii and N. Kirova, Sov. Sci. Rev. Sect. A {\bf 5}, 99
(1984),
\\ {\it Electron Selflocalization and Periodic
               Superstructures in Quasi-One-Dimensional dielectrics}\\
A. J. Heeger, S. Kivelson,
                      J. R. Schrieffer and W.-P. Su,
Rev. Mod. Phys. {\bf 60}, 781-850 (1988).
\\ {\it Solitons in Conducting Polymers}


\bibitem{TJZW-1985}
S. B. Treiman, R. Jackiw, B. Zumino, and E. Witten,
{\it Current Algebra and Anomalies},
Princeton University Press (1985) and references therein.

\bibitem{Diakonov-1997} D. Diakonov, Lectures at the Advanced Summer
 School on Non-Perturbative Field Theory, Peniscola, Spain, June, 1997
(hep-ph/9802298). \\
{\it Chiral
Quark-Soliton Model}

\bibitem{WilczekZee-1983}
F. Wilczek and A. Zee, Phys. Rev. Lett. {\bf 51}, 2250-2252 (1983).
 \\ {\it Linking Numbers, Spin, and Statistics of Solitons}

\bibitem{DFN-1985}
B. A. Dubrovin, A. T. Fomenko, S. P. Novikov,
\\ {\it Modern Geometry-Methods and
Applications : Part II, the Geometry and Topology of Manifolds}
 (Graduate Texts in Mathematics, Vol 104), Springer-Verlag, 1985.

\bibitem{MM-1995}
Y. G. Makhlin and T. S. Misirpashaev,
JETP Lett. {\bf 61}, 49-55 (1995).
\\ {\it Topology of Vortex Soliton
Intersection - Invariants and Torus Homotopy}

\bibitem{Pontrjagin-1941three}
L. Pontrjagin, Recueil Mathematique {\bf 9(51)}, 331-359 (1941).
 \\ {\it A classification of mappings of the three-dimensional complex into the
two-dimensional sphere}

\bibitem{GovShan-1989}
T. R. Govindarajan and R. C. Shankar, Mod. Phys. Lett. A {\bf 4}, 1457-1462
(1989).
\\ {\it Novel Topological Features of the O(3) Non-Linear Sigma Model in 2+1
Dimensions}

\bibitem{Haldane-1983PLA}
F. D. M. Haldane, Phys. Lett. A {\bf 93}, 464-468 (1983). \\ {\it Continuum
 Dynamics of the 1-D Heisenberg
Anti-Ferromagnet-Identification with the O(3) Non-Linear Sigma-Model}

\bibitem{Jaroszewicz-1984} T. Jaroszewicz, Phys. Lett. {\bf B146}, 337-340
(1984).
\\ {\it Induced Fermion Current in
the $\sigma$ Model in (2+1) Dimensions}

\bibitem{ABMNV-1986}
L. Alvarez-Gaume, J. B. Bost, G.
Moore, P. Nelson, and C. Vafa,
Comm. Math. Phys {\bf 112}, 503-552 (1987) and references therein.
\\ {\it Bosonization on Higher
Genus Riemann Surfaces}

\bibitem{Jaroszewicz} T. Jaroszewicz,
Phys. Lett. B {\bf 159}, 299-302 (1985), \\ {\it Induced Topological Terms,
Spin and Statistics in (2+1) Dimensions} \\
T. Jaroszewicz, Phys. Lett. B {\bf 193}, 479-485 (1987). \\ {\it
Fermion-Induced Spin of Solitons: Vacuum and Collective Aspects}

\bibitem{Witten-1983baryons} E. Witten,
Nucl. Phys. {\bf B223}, 433-444 (1983).
\\ {\it Current-Algebra, Baryons, and Quark Confinement}

\bibitem{Tsvelik-1994} A. M. Tsvelik,
Phys. Rev. Lett. {\bf 72}, 1048-1051 (1994). \\ {\it Semiclassical
Solution of One-Dimensional Model of Kondo Insulator}

\bibitem{stability} Homotopy groups
$\pi_{n+k}\,(S^{n})$ are the same for $n\ge k+2$ (sometimes even for
smaller $n$). For
example $\pi_{n}(S^{n})=Z$ for any $n$, $\pi_{n+1}(S^{n})=Z_{2}$ for $n>2$,
but  it is
$Z$ for $n=2$ and $0$ for
$n=1$, etc. See e.g., Ref. \cite{DFN-1985}.

\bibitem{NiemiSemenoff-1986}
For review see:
A. J. Niemi and G. W. Semenoff,
Phys. Rep. {\bf 135}, 99-193 (1986).
\\ {\it Fermion Number Fractionization in Quantum Field Theory}

\bibitem{MKWil-1984ill}
R. MacKenzie and F. Wilczek,
Phys. Rev. D {\bf 30}, 2194-2200 (1984). \\ {\it Illustrations of vacuum
 polarization by solitons}

\bibitem{Witten-1983global} E. Witten,
Nucl. Phys. {\bf B223}, 422-432 (1983).
\\ {\it Global Aspects of Current-Algebra}

\bibitem{BNRS-1982} A. P. Balachandran, V. P. Nair, S. G. Rajeev and A. Stern,
 Phys. Rev. Lett. {\bf 49}, 1124 (1982). \\ {\it Exotic Levels from
Topology in the
Quantum-Chromodynamic Effective Lagrangian}

\bibitem{NiemiSemenoff-1985}
A. J. Niemi and G. W. Semenoff,
Phys. Rev. Lett. {\bf 55}, 927-930 (1985). {\it Quantum Holonomy and the
 Chiral Gauge Anomaly}

\bibitem{Stone-1986}
M. Stone, Phys. Rev. D {\bf 33}, 1191-1194 (1986). \\
{\it Born-Oppenheimer
approximation and the origin of Wess-Zumino terms: Some quantum-mechanical
examples}

\bibitem{PolWieg-1983} A. Polyakov and P. B. Wiegmann, Phys. Lett.
 {\bf B 131}, 121-126 (1983). \\ {\it Theory of Non-Abelian Goldstone
Bosons in 2 Dimensions}

\bibitem{Witten-1984bos}
E. Witten, Commun. Math. Phys. {\bf 92}, 455-472 (1984). \\ {\it Non-Abelian
 Bosonization in Two Dimensions}

\bibitem{Wiegmann-1989} P. B. Wiegmann,
Nucl.Phys. {\bf B323}, 311-329 (1988).
\\ {\it Multivalued Functionals and Geometrical Approach for Quantization of
 Relativistic-Particles and Strings}

\bibitem{KnizhnikZam-1984}
V. G. Knizhnik and A. B. Zamolodchikov,
Nucl. Phys. {\bf B247}, 83-103 (1984).
\\ {\it Current-Algebra and Wess-Zumino
Model in 2 Dimensions}









\end{thebibliography}
\end{document}